\documentstyle[preprint,eqsecnum,aps]{revtex}

\begin{document}
\draft
\title{Skyrmion dynamics on the unstable manifold and 
the nucleon-nucleon interaction}
\author{Th. Waindzoch and J. Wambach}
\address{
Institut f\"ur Kernphysik, Technische Hochschule Darmstadt,\\
Schlo{\ss}gartenstr. 9, D-64289 Darmstadt, Germany}
\date{\today}
\maketitle
\begin{abstract}
The unstable manifold of the $B=2$ sector of the Skyrme model is constructed
numerically using the gradient-flow method. Following paths of steepest 
descent from the $B=2$ hedgehog, we apply a collective coordinate description
for the motion on the manifold to extract a Hamiltonian, approximately valid 
for the quantum description of the low-energy nucleon-nucleon interaction. The 
resulting potential -- obtained in the Born-Oppenheimer approximation -- is 
in qualitative agreement with phenomenological potentials. 
\end{abstract}

\narrowtext
Because of the non-abelian character of $QCD$, a first-principles derivation of 
low-energy hadronic physics has proven exceedingly difficult.
Based on the large-$N_c$ limit, where $N_c$ is the number of colors
\cite{thooft72,witten79} an interesting non-perturbative approach is offered
by effective chiral meson models, as witnessed by their great
phenomenological success. A popular realization is the Skyrme model 
-- rather simple in nature yet remarkably realistic \cite{skyrme62}. 
Its virtue lies in the fact that the non-linear classical field 
equations allow for topological solitonic solutions, called skyrmions, which 
carry baryon number, $B$ \cite{witten83}, thus unifying the mesonic and 
baryonic sector in a common description. Aside from single-baryon properties, 
the multi-baryon problem is of great significance in an attempt to understand
complex nuclei within this framework. This poses a difficult problem
since the starting point is inherently classical and 'requantization' is highly
non-trivial. Most promising, in this respect, is the $B=2$ sector,
especially the description of nucleon-nucleon scattering, which will be the
focus of the present work.

The Skyrme model is specified by the Lagrange density 
\begin{equation}
{\cal L}= {{f_\pi^2}\over 4} Tr[{\partial_\mu U \partial^\mu U^\dagger}]
+ {1\over{32 e_s^2}}Tr([U^\dagger \partial_\mu U,U^\dagger \partial_\nu U])^2 
+ {{m_\pi^2 f_\pi^2}\over 2} Tr[U-1] ,\eqnum{1}
\end{equation}
where $U\epsilon SU(2)$ denotes the non-linear pion field. The first term
is the familiar non-linear sigma model, while the second is one of the possible
four-derivative terms, derived systematically in chiral perturbation theory.
A representation of the $U$-matrix is of the form 
$U=(\sigma + i\vec{\tau}\vec{\pi})/f_\pi$, with $\vec{\tau}$ being the three
Pauli matrices. The parameters of the model are the pion decay 
constant $f_\pi$ (=93 MeV), the Skyrme parameter $e_s$ (=4.76) , related to the
$\rho\pi\pi$-coupling constant \cite{bhadu} and the pion mass $m_\pi$ 
(=138 MeV). The values quoted in brackets, which will be used in our
calculations, ensure the correct asymptotic behavior of the fields, 
rendering the correct one-pion exchange tail of the nucleon-nucleon potential 
\cite{walwam92}. 

In order to assess the applicability of the Skyrme model beyond the
$B=1$ sector, where one gets reasonable results for the properties of the 
nucleon \cite{adkins83}, a study of the $B=2$ sector should yield an acceptable 
description of nucleon-nucleon scattering as well as the deuteron bound-state.
Starting from classical skyrmion-skyrmion interactions one has to decide on
the appropriate field configurations in the $B=2$ sector. 
Since the Skyrme-Lagrangian (1) is viewed as a model for large-$N_c$ $QCD$, 
the semi-classical quantization procedure demands that we start from the 
minimal-energy configurations. As has been argued by Manton \cite{manton88}, 
the appropriate procedure to exclude undesirable vibrational modes, is to 
calculate the classical fields on the so-called 'unstable manifold' of  
$B=2$ configurations. In this respect, an important role is played by the
$B=2$ 'hedgehog', which is of the form $U=e^{i\vec{\tau}\vec{\pi}F(r)}$,
characterized by the 'profile function' $F(r)$ with boundary conditions:
$F(0)=2\pi$ and $F(\infty)=0$. This special configuration is a saddle point on 
the hyperplane of minimal energy configurations and can therefore serve 
as a 'source' for further, less energetic $B=2$ configurations. 
A complete class of configurations can be obtained numerically via the 
gradient-flow method \cite{atiyah93}. One evolves the fields along a
path of steepest descent, originating from the B=2 hedgehog which is
initially excited by an unstable mode. The set of all decay paths defines a 
twelve-dimensional unstable manifold, called $M^{12}$, to be identified as 
the manifold of minimal-energy configurations in the $B=2$ sector. The 
unstable modes of the $B=2$ hedgehog have been reported in \cite{wirzba90} by
treating the hedgehog on compactified spatial 'hypersphere' and have been 
recalculated in \cite{waiwam96} in 'flat' space, using lattice discretization.
In the latter work the gradient-flow method and its numerical implementation
was also developed in detail. 

The calculations in \cite{waiwam96} were performed on a three-dimensional grid 
using a finite differencing scheme for solving the gradient-flow equation. 
Applying a collective-coordinate description to the motion on $M^{12}$, which 
is always possible locally, the gradient-flow equation in collective 
coordinate space reads as
\begin{equation}
M_{ij}(Q){{\partial Q_j}\over {\partial \tilde{t}}}= 
- {{\partial V(Q)}\over {\partial Q_i}}  \label{gradflow}.\eqnum{2}
\end{equation}
The evolution parameter $\tilde{t}$ is a pseudo-time, having a dimension
$($time$)^2$. The matrix $M_{ij}(Q)$ is the metric- or mass tensor on $M^{12}$,
while $V(Q)$ denotes the interaction potential between two skyrmions. Both 
quantities depend, in general, on three relative collective coordinates of the 
$B=2 $ system. For well-separated skyrmions, these coordinates can be defined 
as the distance $R$ and the relative isospin orientation $C$ between the 
skyrmions. Taking the separation axis -- in the body-fixed frame -- along the 
z-direction, the matrix $C$ can be written in terms of two Euler angles 
$\beta$ and $\chi$ as \cite{walwam92}
\begin{equation}
C=\cos({\beta\over 2})\cos({\chi\over 2})+i\tau_2\sin({\beta\over 2})
+i\tau_3 \cos({\beta\over 2}) \sin({\chi\over 2}) .\eqnum{3}
\end{equation}
In addition to the three relative coordinates, a $B=2$ configuration
is characterized by nine global collective coordinates, which fix the
position of the center of mass and the overall orientation in space and 
isospace. Subsequently, we shall treat these coordinates in the adiabatic 
approximation, since they describe the 'zero-mode motion' of the skyrmions.

By exciting the $B=2$ hedgehog with different linear combinations of the 
elementary unstable modes, one can -- in principle -- construct the entire 
unstable manifold numerically. Denoting the 
six unstable modes of the $B=2$ hedgehog by $\delta U^{(M,i)}$ (magnetic mode
\cite{wirzba90}) and $\delta U^{(E,i)}$ (electric mode \cite{wirzba90}), where 
$i=x,y,z$ specifies the symmetry axis of the 
unstable mode (for details see \cite{waiwam96}), a general combination of the 
unstable modes, in the body-fixed frame, can be written as
\begin{equation}
\delta U = (\cos \varphi)\delta U^{(M,z)} + (\sin \varphi)\left[(\cos \theta)
\delta U^{(E,z)}+(\sin \theta) \delta U^{(E,x)}\right] .\eqnum{4} \label{fluct}
\end{equation}
All other combinations of the unstable modes are related to one of the 
combinations above plus a global transformation of the $B=2$ hedgehog. Thus, 
in the body-fixed frame, only the two mixing angles
$\varphi$ and $\theta$ are of importance. Together with the pseudo-time 
parameter $\tilde{t}$, these are three collective coordinates which uniquely 
specify the field configurations along the paths of steepest descent. As 
discussed above, another complete set of collective coordinates is given by the 
three relative collective coordinates $R$ and $C$. 
In numerical calculations the latter are easily extractable only for 
well-separated skyrmions.

Using the finite-difference method as in \cite{waiwam96}, we calculate the 
paths of steepest descent from the $B=2$ hedgehog numerically. 
Several paths are shown in Fig.~1 and Fig.~ 2, where 
the interaction potential along a given path is displayed
as a function of the distance $R$ between two skyrmions.
For the distance coordinate we use a quadrupole definition \cite{waiwam96}.
The computational details can be found in \cite{wai97}.
Depending on the initial mixing angles in (\ref{fluct}), the excited $B=2$ 
hedgehog, positioned at the maximum of the interaction potential, can 
either split into two distinct $B=1$ skyrmions (Fig.~1), or the paths end in 
the field configuration of minimal energy -- the well known torus 
configuration (Fig.~2). The latter is classically bound by 120 MeV and is a 
source for large attraction in the interaction potential \cite{walwam92}. 
The path of steepest descent which results from the
gradient-flow of two well-separated attractive skyrmions ($\beta=\pi$) is also 
shown in Fig.~2. The field configurations along this
path can be described entirely by the relative distance coordinate which, 
together with the nine global collective coordinates, lead to a 
ten-dimensional submanifold of $M^{12}$, usually called $M^{10}$ 
\cite{manton88,verbaar87,manton95}. 

Figs.~1 and 2 indicate that the motion on the unstable manifold is
generally quite complicated. The skyrmions move relative to each other in $R$ 
and also tend to rotate in $C$ whenever necessary in order to approach a 
configuration with less energy. As discussed in \cite{waiwam96}, we can 
identify the relative isospin orientation $C$
between skyrmions for three different paths of steepest descent on $M^{12}$. 
Along the two elementary decay paths from the $B=2$ hedgehog for excitations 
with the unstable modes superimposed by mixing angles of $(\varphi=0)$ and 
$(\varphi={\pi\over 2},\theta=0)$, the skyrmions have 
fixed relative isospin orientations $C=1$ and $C=i\tau_3$, respectively. 
Along the gradient-flow path originating from two-well separated 
attractive skyrmions, the $B=2$ configurations retain their relative isospin 
orientation, $C=i\tau_2$. 
Fig.~ 3 shows these three important paths on the unstable manifold $M^{12}$. 
Performing a finite expansion of the potential energy in terms of 
Wigner-D-functions, as explained in \cite{walwam92}, 
the skyrmion-skyrmion potential can be written approximately as 
\begin{equation}
V(R,\beta,\chi)= V_{00}(R)+V_{10}(R)\cos(\beta) 
+ V_{11}(R)\cos^2({\beta\over 2})\cos(\chi),\eqnum{5}
\label{pot}
\end{equation}
where the radial coefficients $V_{00}, V_{10}$ and $V_{11}$ are deduced from 
the three potentials shown in Fig.~3. 
We expect that the expansion will hold for medium and large distances. 
At shorter distances, higher-order terms in the expansion, 
may become important, however. 

Inspection of the element $M_{RR}(R,C)$ in the reduced mass tensor along the 
three important paths on $M^{12}$, using the method of \cite{waiwam96}, 
shows that this quantity depends rather strongly on the collective coordinates 
at shorter distances. In order to keep the reduced mass fixed at its asymptotic 
value $\mu_s=M_s/2$, where $M_s=1463$ MeV is the single-skyrmion mass in our 
parameter set, we change the definition of the distance R on $M^{12}$. 
Different choices are possible since, a priori, no distance definition is 
preferred. The only constraint is that all definitions lead to the same result 
for well separated skyrmions. For the three relevant paths the new distance 
can be easily calculated \cite{wai97}. The results for the interaction
potential are also shown in Fig.~3. Due to the redefinition of the distance 
the potential (\ref{pot}) is modified at shorter distances. This indicates 
further that a detailed knowledge of the metric structure on $M^{12}$ may 
become of importance for a proper description.

Since a more thorough calculation of the mass tensor $M_{ij}(Q)$ on $M^{12}$ 
is beyond our means at present, we use its large distance form, in order 
to define a collective coordinate representation of the dynamics. The 
approximate Lagrange function in the body-fixed frame then becomes  
\begin{equation}
L^{b.f.}= {{\mu_s}\over 2}\dot{R}^2 + {\lambda_s\over 2}(\dot{\beta}^2 
+ \cos^2({\beta\over 2})
\dot{\chi}^2)-V(R,\beta,\chi) .\eqnum{6} \label{lag}
\end{equation}
The quantity $\lambda_s=(265 MeV)^{-1}$ is the moment of inertia of a $B=1$ 
skyrmion in our parameter set. The quality of the approximation can be
tested by solving the corresponding gradient-flow equations for different 
initial values of $R,\beta$ and $\chi$. As a result, we find qualitative 
agreement of the resulting flow curves with paths of steepest descent on the 
unstable manifold. Unfortunately, a precise comparison is not possible at the 
moment, because of the difficulty of a rigorous definition of the 
relative collective coordinates for all paths on $M^{12}$.

Treating the nine global collective coordinates in an adiabatic approximation 
and using the large-distance mass tensor on $M^{12}$, we 
arrive -- after canonical quantization of all collective coordinates --
at a simple Hamiltonian for the $B=2$ sector. In the c.m. system it reads
\begin{equation}
H={\hat{p}_{R}^2\over {2\mu_s}}+{\vec{L}^2\over{2\mu_s R^2}}
+{1\over {2\lambda_s}}(\vec{S}_1^2+\vec{S}_2^2)+2M_s+V(R,C) .\eqnum{7}
\end{equation}
Here $\hat{p}_R$ denotes the radial momentum operator, $\vec{L}$ is the 
angular momentum operator and $\vec{S}_{1,2}$ are the spin operators of the 
two interacting baryons. As has been done earlier, we apply a Born-Oppenheimer 
approximation to the collective coordinate dynamics, in order to deduce a 
nucleon-nucleon potential from the Hamiltonian above
\cite{faessler86}. Taking into account finite-$N_c$ effects \cite{walwam92}, 
the reduced Hamiltonian is diagonalized for given $R $ in the space of 
asymptotic nucleon and $\Delta$-isobar states. This leads to a 
nucleon-nucleon potential of the form
\begin{eqnarray}
V_{NN}^{(BO)}(R)&=& V_c^0(R)+V_c^1(R)(\vec{\tau}_1\vec{\tau}_2)
+V_\sigma^0(R)(\vec{\sigma}_1\vec{\sigma}_2)
+V_\sigma^1(R)(\vec{\sigma}_1\vec{\sigma}_2)(\vec{\tau}_1\vec{\tau}_2)
\nonumber \\
&&+V_{T}^0(R)S_{12}+V_{T}^1(R)S_{12}(\vec{\tau}_1\vec{\tau}_2) ,\eqnum{8}
\end{eqnarray}
where $S_{12}$ denotes the tensor operator.
Fig.~ 4 shows as an example the central part $V_c^0(R)$ in comparison to the 
phenomenologically successful Argonne potential \cite{wiringa84}. 
The potential deduced from our nuclear Hamiltonian renders almost all
features of the realistic potential. It has the correct asymptotic strength,  
reproduces a medium-range attraction and builds up a repulsive core at 
shorter distances. This demonstrates that the Skyrme model can provide sizable
attraction between nucleons and therefore might serve as a good model for the 
study of bound states. 

In summary, we have presented the first calculation of field configurations 
on the unstable manifold in the $B=2$ sector of the Skyrme model. Applying a 
collective coordinate description, we were able to define a simple nuclear 
Hamiltonian, approximately valid at medium and large distances. 
When performing a calculation of the nucleon-nucleon potential within the
Born-Oppenheimer approximation, good qualitatively agreement with
phenomenological potentials for the central part has been found. To improve on 
the results, one has to examine the metric structure on the unstable manifold 
$M^{12}$, in detail. It has been demonstrated that such a analysis is 
numerically possible. 
\acknowledgments
We would like to thank N.S. Manton for fruitful discussion.

\begin{figure}[h]
\vskip 8.0cm
\includegraphics{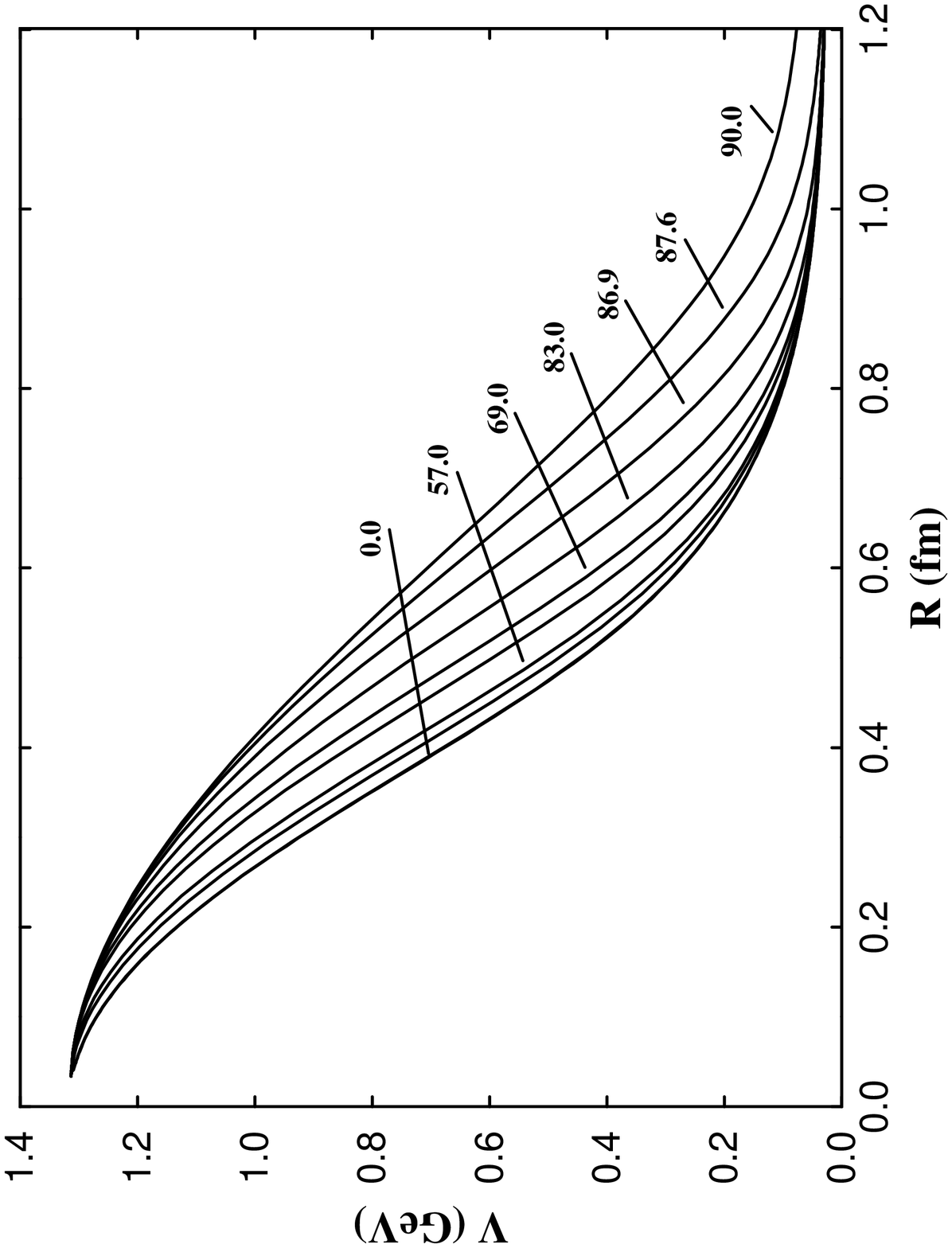}
\caption{The interaction potential of two skyrmions as a function of the 
quadrupole distance $R$ along paths of steepest descent from the $B=2$ 
hedgehog. Shown are paths resulting from a 'parallel superposition' of 
the unstable modes $(\theta=0)$, using different mixing angles $\varphi$ 
(in degree).}
\end{figure}

\begin{figure}[h]
\vskip 8.0cm
\includegraphics{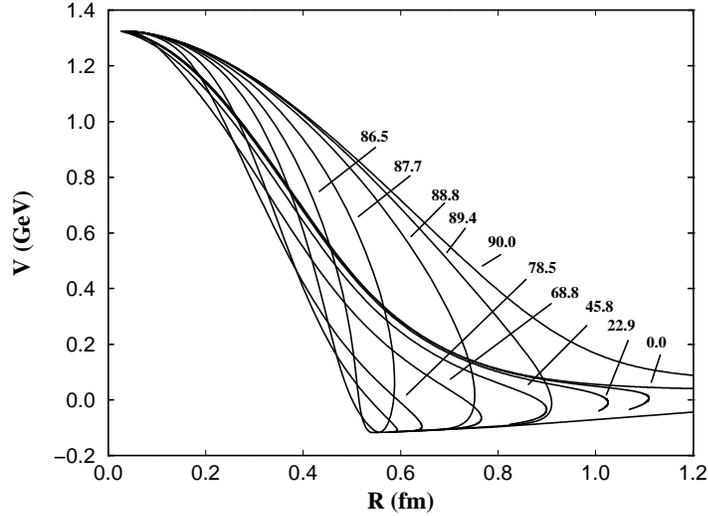}
\caption{A different set of paths of steepest descent from the $B=2$ hedgehog.
Shown are paths resulting from an 'orthogonal superposition' of the unstable 
modes $(\theta=90$ degree$)$, using different mixing angles $\varphi$ 
(in degree).}
\end{figure}

\newpage
$ $

\begin{figure}[h]
\vskip 8.0cm
\includegraphics{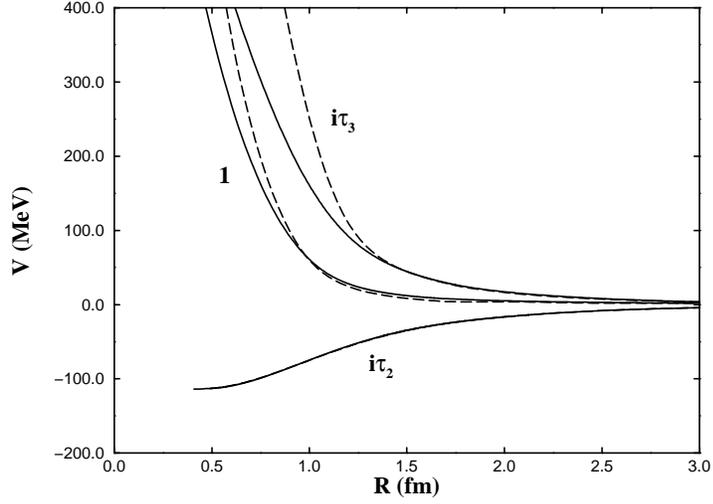}
\caption{The skyrmion-skyrmion potential as a function of the distance $R$
and the relative isospin orientation $C$. The dashed lines give the results 
using a quadrupole definition for the distance while the solid lines denote the 
results using a redefined distance definition, as explained in the text.}
\end{figure}

\begin{figure}[b]
\vskip 8.0cm
\includegraphics{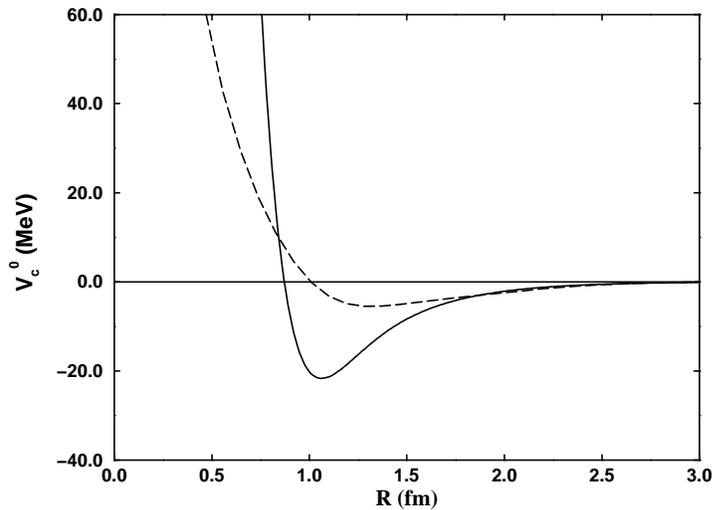}
\caption{The central part of the nucleon-nucleon potential for the 
Argonne potential (solid line) and for the Skyrme model, using the results from 
the unstable manifold and the Born-Oppenheimer approximation (dashed line).}
\end{figure}

\end{document}